\documentclass[aps,prc,twocolumn,nofootinbib]{revtex4-1}
\usepackage[utf8]{inputenc}
\usepackage{csquotes}
\usepackage{todonotes}
\usepackage{amsmath,amssymb}
\usepackage{physics}
\usepackage{units}
\usepackage{currfile}
\usepackage{bbold}
\usepackage{textcomp}

\usepackage{hyperref}

\DeclareMathOperator*{\argmax}{argmax}

\newcommand{\chieft}{$\chi$EFT}
\newcommand{\NN}{$NN$}
\newcommand{\NNN}{$NNN$}
\newcommand{\nn}{$nn$}
\newcommand{\np}{$np$}
\newcommand{\pp}{$pp$}
\newcommand{\piN}{$\pi N$}
\newcommand{\Ctnn}{
	\ifmmode \widetilde{C}_{1S0}^{nn} \else$\widetilde{C}_{1S0}^{nn}$\fi
}
\newcommand{\Ctnp}{
	\ifmmode \widetilde{C}_{1S0}^{np} \else$\widetilde{C}_{1S0}^{np}$\fi
}
\newcommand{\Ctpp}{
	\ifmmode \widetilde{C}_{1S0}^{pp} \else$\widetilde{C}_{1S0}^{pp}$\fi
}
\newcommand{\prob}{\textnormal{pr}}
\newcommand{\lec}{\alpha}
\newcommand{\lecnn}{\alpha_{nn}}
\newcommand{\pdf}{PDF}
\newcommand{\ppd}{PPD}
\newcommand{\lecs}{\ensuremath{\vec{\lec}}}
\newcommand{\hmclecs}{\ensuremath{\vec{\lec}_{np,pp}}}
\newcommand{\Ct}{\widetilde{C}}
\newcommand{\hmcdata}{D_{np,pp}}
\newcommand{\subnn}{_{nn}}
\newcommand{\supnn}{^{nn}}
\newcommand{\subnp}{_{np}}
\newcommand{\supnp}{^{np}}
\newcommand{\subpp}{_{pp}}
\newcommand{\suppp}{^{pp}}

\newcommand{\subth}{_\textnormal{th}}
\newcommand{\subexp}{_\textnormal{exp}}
\newcommand{\subref}{_\textnormal{ref}}

\newcommand{\aexpnn}{a\subexp\supnn}
\newcommand{\athnn}{a\subth\supnn}
\newcommand{\athnp}{a\subth\supnp}
\newcommand{\athpp}{a\subth\suppp}
\newcommand{\normal}{\mathcal{N}}

\newcommand{\invgamma}{\mathcal{IG}}
\newcommand{\cbar}{\bar{c}}
\newcommand{\hmccbar}{\ensuremath{\bar{c}_{np,pp}}}
\newcommand{\cbarsq}{\cbar^2}
\newcommand{\hmccbarsq}{\ensuremath{\hmccbar^2}}
\newcommand{\vecc}{\vec{c}}
\newcommand{\veccsq}{\vec{c}^{\,2}}

\newcommand{\Ctunit}{
  \ifmmode \times 10^4\,\text{GeV}^{-2} \else $\times 10^4\,$GeV$^{-2}$ \fi
}
\newcommand{\Cunit}{
  \ifmmode \times 10^4\,\text{GeV}^{-4} \else $\times 10^4\,$GeV$^{-4}$ \fi
}
\newcommand{\Dunit}{
  \ifmmode \times 10^4\,\text{GeV}^{-6} \else $\times 10^4\,$GeV$^{-6}$ \fi
}
\newcommand{\cunit}{
  \ifmmode ~\text{GeV}^{-1} \else GeV$^{-1}$ \fi
}
\newcommand{\Tlab}{
  \ifmmode T_{\textnormal{lab}} \else $T_{\textnormal{lab}}$ \fi
}
\newcommand{\lo}{
  \ifmmode \textnormal{LO} \else\unskip LO\fi
}
\newcommand{\nlo}{
  \ifmmode \textnormal{NLO} \else\unskip NLO\fi
}
\newcommand{\nnlo}{
  \ifmmode \textnormal{NNLO} \else\unskip NNLO\fi
}
\newcommand{\nnnlo}{
  \ifmmode \textnormal{N3LO} \else\unskip N3LO\fi
}

\newcommand{\hmcpr}{\ensuremath{\prob\left(\hmclecs | \hmcdata, \hmccbarsq, I \right)}}
\newcommand{\fullpr}[1]{\ensuremath{\prob\left(\lecs_{#1}, \cbarsq_{#1} | D, \vecc, I \right)}}

\newcommand{\cnew}{c_+}
\newcommand{\IGa}{\alpha}
\newcommand{\IGb}{\beta}
\newcommand{\IGanew}{\IGa_+}
\newcommand{\IGbnew}{\IGb_+}
\newcommand{\IGaprior}{\IGa_0}
\newcommand{\IGbprior}{\IGb_0}

\begin{document}

\title{Bayesian estimation of the low-energy constants up to fourth order in the nucleon-nucleon sector of chiral effective field theory}

\date{\today}

\author{Isak Svensson}
\email{isak.svensson@chalmers.se}
\affiliation{Department of Physics, Chalmers University of Technology, SE-412 96 G\"oteborg, Sweden}

\author{Andreas Ekstr\"om}
\affiliation{Department of Physics, Chalmers University of Technology, SE-412 96 G\"oteborg, Sweden}

\author{Christian Forss\'en}
\affiliation{Department of Physics, Chalmers University of Technology, SE-412 96 G\"oteborg, Sweden}

\begin{abstract}
We use Bayesian methods and Hamiltonian Monte Carlo (HMC) sampling to infer the posterior probability density function (\pdf) for the low-energy constants (LECs) up to next-to-next-to-next-to-leading order (\nnnlo) in a chiral effective field theory (\chieft{}) description of the nucleon-nucleon interaction. In a first step, we condition the inference on neutron-proton and proton-proton scattering data and account for uncorrelated \chieft{} truncation errors. We demonstrate how to successfully sample the 31-dimensional space of LECs at \nnnlo\ using a revised HMC inference protocol. In a second step we extend the analysis by means of importance sampling and an empirical determination of the neutron-neutron scattering length to infer the posterior \pdf{} for the leading charge-dependent contact LEC in the $^{1}S_0$ neutron-neutron interaction channel. While doing so we account for the \chieft{} truncation error via a conjugate prior. We use the resulting posterior \pdf{} to sample the posterior predictive distributions for the effective range parameters in the $^{1}S_0$ wave as well as the strengths of charge-symmetry breaking and charge-independence breaking. We conclude that empirical point-estimate results of isospin breaking in the $^{1}S_0$ channel are consistent with the \pdf s obtained in our Bayesian analysis and that, when accounting for \chieft{} truncation errors, one must go to next-to-next-to-leading order to confidently detect isospin breaking effects.
\end{abstract}

\maketitle

\section{Introduction}
\label{sec:intro}
In chiral effective field theory (\chieft)~\cite{Weinberg:1990rz,Epelbaum:2008ga,Machleidt:2011zz,Hammer:2019poc} the nuclear interaction is parametrized in terms of low-energy constants (LECs) that capture unresolved physics and must be determined from data. The number of LECs grows with the order of the chiral expansion, and the parameter estimation becomes a challenging inference problem that is directly connected with the precision of the theory.
Moreover, in \chieft{} the long-range pion-nucleon (\piN) interaction appears as a subprocess of the nuclear interaction. We can therefore constrain the long-range part of the nuclear interaction rather well using measured \piN{} scattering data~\cite{Hoferichter:2015hva}, albeit less so in the delta-full sector of \chieft{}~\cite{Siemens:2016jwj}. In a Bayesian context, this knowledge can be straightforwardly accounted for as a prior when learning more about the nuclear interaction from new data, as demonstrated in, e.g., Refs.~\cite{Wesolowski:2021cni,Svensson:2021lzs}. In fact, Bayesian inference methods allow us to account for any prior beliefs about \chieft, most importantly its truncation error~\cite{Furnstahl:2015rha}. This highlights some of the central synergies of \chieft{} and Bayesian methods. In this paper, we focus on two challenges tied to a Bayesian analysis of the nucleon-nucleon (\NN) interaction:
(i) how to reliably sample the high-dimensional Bayesian posterior probability density functions (\pdf s) of the LECs up to next-to-next-to-next-to-leading order (\nnnlo) in \chieft,
and (ii) how to extend these posterior \pdf s, here inferred from neutron-proton (\np) and proton-proton (\pp) scattering data, to account for the uncertainty in the isospin breaking (IB) and leading neutron-neutron (\nn) short-range LEC acting in the $^{1}S_0$ partial wave.
Throughout the paper we will often use the short-hand labels \emph{posterior} and \emph{prior} to indicate posterior and prior \pdf s, respectively.

Sampling a high-dimensional \pdf{} poses a significant challenge in any Bayesian analysis. In a previous paper~\cite{Svensson:2021lzs} we explicitly demonstrated the efficiency and manageable dimensional scaling of the Hamiltonian Monte Carlo (HMC)~\cite{duane87} algorithm applied to \chieft{} up to next-to-next-to-leading order (\nnlo). HMC exploits the geometry of the parameter space and Hamiltonian dynamics to draw virtually independent Markov chain Monte Carlo (MCMC) samples concentrated to the bulk of the probability mass. In this paper, we revise our HMC protocol to sample the LEC posterior at the next chiral order, i.e., \nnnlo, where we encounter a significantly more challenging inferential problem in 31 dimensions, one per LEC.

Furthermore, the fundamental effect of IB leads to slight differences in the strong interaction between neutrons, protons, and between protons and neutrons. It originates from differences in the masses and electromagnetic charges of the up- and down-quarks~\cite{Miller:1990iz} and is expected to be weaker in the irreducible interaction between three nucleons~\cite{Friar:2004ca,Epelbaum:2004xf}. Although weak, the IB of the strong nuclear interaction plays an important role in \textit{ab initio}, and mean-field, analyses of infinite nuclear matter and finite nuclei, in particular towards the driplines with pronounced proton-to-neutron ratios (see, e.g., Refs.~\cite{Hagen:2015yea,GarciaRuiz:2016ohj,Brown:2017xxo,Koszorus:2020mgn,Roca-Maza:2018bpv,Hu:2021trw,Novario:2021low,Naito:2021uyk,Naito:2022hyb,Reinhard:2022jby}). It is therefore important to quantify the uncertainties pertaining to the LECs governing the strengths of the IB effects. Unfortunately, inferences conditioned solely on the world database of \NN{} scattering data~\cite{perez13-1,perez13-2}, which does not contain \nn{} cross sections, leaves the posterior of LECs acting (only) in the \nn{} isospin channel unconstrained. To handle this, a point estimate of the leading charge-dependent LEC in the \nn{} channel is typically obtained using the empirical value for the corresponding scattering length, primarily in the $^{1}S_0$ partial wave. This latter quantity parametrizes the total \nn{} cross section in the limit of zero scattering energy and its value is estimated from data on hadronic reactions that involve two neutrons in the initial and/or final state~\cite{Gardestig:2009ya}. A future prospect is to employ results from realistic lattice quantum chromodynamics to directly infer the values of the LECs~\cite{Chang:2018lqcd}.

In \chieft, isospin is an exact symmetry at leading order (\lo)~\cite{vanKolck:1995cb}. At next-to-leading order (\nlo), one introduces realistic charged-to-neutral pion mass splittings in the one-pion exchange potential (OPEP). The $\approx 3\%$ mass splitting of the rather light pions induces IB in the $S$ waves and beyond. We also have charge-dependent and nonderivative LECs in the $^{1}S_0$ partial wave. Higher-order IB effects can be accounted for systematically by introducing, e.g., charge-dependent \piN{} LECs, (charged) pion-photon interactions, and charge-dependent LECs in partial waves with nonzero angular momentum. However, many of those IB effects are estimated to be negligible compared to the OPEP mass splitting and nonderivative $S$-wave LECs~\cite{Epelbaum:2008ga,Machleidt:2011zz,Hammer:2019poc}. Indeed, the leading IB \piN\ LECs were recently~\cite{Reinert:2020mcu} inferred using \NN{} scattering data and fifth-order \chieft{}, i.e., N4LO, and found to exhibit no significant charge dependence. In fact, higher-order IB effects are often neglected in quantitative chiral interactions~\cite{Piarulli:2014bda,Ekstrom:2015rta,Reinert:2017usi,Epelbaum:2014sza,Entem:2017gor}.

In this paper, we employ \NN{} interactions from \chieft{} up to \nnnlo\ in Weinberg power counting as defined in Ref.~\cite{Machleidt:2011zz}, and with the IB effects due to pion mass-splitting in the OPEP and charge-dependent leading $S$-wave contacts. We use nonlocal regulators in relative momenta $p$ according to $f(p) = \exp(-p^{2n}/\Lambda^{2n})$ with a cutoff $\Lambda=450$ MeV and $n=3$. Two-pion exchanges are spectral-function regulated~\cite{Epelbaum:2003gr,Epelbaum:2003xx} with a cutoff of 700 MeV. At \nnnlo, to complete the subleading two-pion exchange, we include all two-loop diagrams with some of them evaluated using numerical integration. To remedy the on-shell redundancy~\cite{Reinert:2017usi,Wesolowski:2018lzj} in the $S$-wave contact potential at \nnnlo\ we set the contact fourth-order contact LECs $\widehat{D}_{^1S_0}$, $\widehat{D}_{^3S_1}$, and $\widehat{D}_{^3S_1 - ^3D_1}$ (in the notation of Ref.~\cite{Machleidt:2011zz}) to zero. For describing \pp\ and \np\ low-energy scattering data we append the standard electromagnetic interactions up to second order in the fine-structure constant as outlined in, e.g., Ref.~\cite{Carlsson:2015vda}, at all chiral orders.

Within this \chieft{} framework, we perform a Bayesian study of the \NN{} interaction, up to N3LO, conditioned on \NN{} scattering data and subsequently extend the LEC posterior using an empirical value for the $^{1}S_0$ \nn\ scattering length $\aexpnn$. This allows us to quantify the uncertainties of the IB effects due to the short-range LECs in \chieft{}. In the process of doing so, we also give an example of the flexibility of the Bayesian framework to straightforwardly expand existing results by introducing new parameters and conditioning on new data. We also test the robustness of a commonly employed model~\cite{Furnstahl:2015rha,Wesolowski:2018lzj} for estimating truncation errors in \chieft{}.

This paper is organized as follows. In Sec.~\ref{sec:stat_meth}, we outline the statistical model upon which we base all inferences and demonstrate how to draw samples from the posterior \pdf s and analyze the consistency of our inferences. In Sec.~\ref{sec:results}, we draw samples from the posterior predictive distributions (\ppd s) for scattering lengths and effective ranges in the $^{1}S_0$ partial wave. We summarize our findings in Sec.~\ref{sec:summary}.

\section{Statistical method}
\label{sec:stat_meth}
In this section we explain our method for inferring the joint posterior \pdf s for \textit{all} the LECs in the \NN\ sector of \chieft\ up to N3LO. After the specification of the prior and likelihood, our inference is performed in two stages. In a first step, we apply the HMC algorithm from Ref.~\cite{Svensson:2021lzs} to quantify the posterior \pdf s for the \NN\ LECs in the \np\ and \pp\ sectors conditioned on \np\ and \pp\ scattering data. The posteriors presented in this paper account for an uncorrelated \chieft{} error model conditioned on the order-by-order convergence pattern up to N3LO. In a second step, we marginalize-in the \nn\ non-derivative LEC \Ctnn\ and condition the inference on a single datum: the \nn\ scattering length in the $^{1}S_0$ partial wave, $\aexpnn = -18.9 \pm 0.4$ fm~\cite{Machleidt:2001rw,Gardestig:2009ya}. We note that this is one of the currently accepted values for this scattering length and that there are conflicting experimental values for which the experimental uncertainties are not fully understood. However, we do not account for this additional level of uncertainty. See, e.g., Ref.~\cite{Gobel:2021pvw} for a summary of the present status on this topic and the proposal of a novel method to measure the \nn{} scattering length. In this paper we only operate with empirical scattering lengths and effective ranges for which electromagnetic (EM) effects have been removed by the originators of those values. Similarly, our theoretical predictions for effective range parameters do not include any EM effects either. We do however include EM effects when we compute scattering cross sections during the HMC sampling.

\subsection{Finding an expression for the joint posterior \label{sec:setting_up_the_inference}}
Using the notation $\prob(A | B)$ for the conditional probability that proposition $A$ is true given $B$, the posterior \pdf\ of interest $\prob(\lecs | D, I)$ can, according to Bayes' theorem, be evaluated as
\begin{equation}
\label{eq:bayes}
\prob(\lecs|D,I) \propto \prob(D|\lecs,I) \times \prob(\lecs|I),
\end{equation}
i.e., as a product of the data likelihood $\prob(D|\lecs,I)$ times the prior \pdf{} $\prob(\lecs|I)$. Here, $\lecs$ denotes the vector of LECs to be inferred, $D$ is employed data, and $I$ encompasses all other assumptions and given information. The given information includes, for example, the chiral order, the regulator cutoff, masses, etc., and we make assumptions about the truncation error, data selection, and so on. Throughout this paper, we omit the overall normalization factor $\prob(D|I)$ as it does not play a central role in parameter estimation.

In a first step we use the \np\ and \pp\ scattering data $\hmcdata$ contained in the Granada database~\cite{perez13-1,perez13-2}.  In doing so we obtained posteriors for the \piN\ and contact \NN\ LECs, excluding the \Ctnn\ LEC. To simplify the notation we often denote this latter, explicitly charge-dependent, \nn\ LEC as $\lecnn$ while all other LECs are collectively denoted $\hmclecs$. In this notation, the main objective in this paper is to extract the joint posterior $\prob(\lecs | D, I)$ of $\lecs \equiv (\lecnn,\hmclecs)$ conditional on $D = (\aexpnn, \hmcdata)$.

Using the product rule of probabilities we can write
\begin{widetext}
\begin{equation}
\begin{split}
\label{eq:lec_posterior}
\prob(\lecs| D, I) & \equiv \prob(\lecnn,\hmclecs | \aexpnn, \hmcdata, I)
\prob\left(\lecnn | \hmclecs, \aexpnn, \hmcdata, I\right) \times \prob(\hmclecs | \aexpnn, \hmcdata, I) \\
&= \prob\left(\lecnn | \hmclecs, \aexpnn, I \right) \times \prob\left(\hmclecs | \hmcdata, I\right),
\end{split}
\end{equation}
\end{widetext}
where we have assumed conditional independence in the final equality.
Thus far, no strong assumptions have been made regarding the relation between $\lecnn$ and $\hmclecs$, and the analysis is quite general.

We use HMC to sample $\prob(\hmclecs | \hmcdata, I)$, and the strategy that we employ for this is explained in Ref.~\cite{Svensson:2021lzs}. In brief, we employ order-by-order differences up to \nnnlo, omitting the \lo\ results, to estimate the variance of the (uncorrelated) \chieft{} truncation error for describing scattering data. We place a multivariate Gaussian prior on the subleading \piN{} LECs at \nnlo\ and \nnnlo\ using the results from a Roy-Steiner analysis of \piN{} scattering amplitudes~\cite{Siemens:2016jwj}. Furthermore, we place a rather weak prior on the \NN{} contact LECs at all orders using an uncorrelated Gaussian \pdf{} with zero mean and standard deviation of $5 \times 10^{4}$ GeV$^{-(k+2)}$ for the LECs belonging to the $k=0,2,4$ (\lo, \nlo, \nnnlo) contact Lagrangian as defined in Weinberg power counting. The full prior factorizes into independent \NN\ and \piN\ \pdf s since we assume no correlation between these sectors. In Sec.~\ref{sec:sampling_using_hmc} we present further details about the sampling and how we revised the sampling protocol to reach \nnnlo. In the remainder of this section we assume that $\prob(\hmclecs | \hmcdata, I)$ is known to us.

Next, we extend this posterior by incorporating $\lecnn$. Using Bayes' theorem, we rewrite the first factor in the final row of~\eqref{eq:lec_posterior} according to
\begin{equation}
\begin{split}
\prob(\lecnn | \hmclecs, \aexpnn, I) \propto&\ \prob(\aexpnn | \lecnn, \hmclecs, I) \\
&\times \prob(\lecnn |I),
\end{split}
\end{equation}
where we again used that the prior for $\lecnn$ is conditionally independent of $\hmclecs$. To evaluate the likelihood $\prob(\aexpnn | \lecnn, \hmclecs, I)$ we numerically compute the scattering length at the specified chiral order in \chieft\ and given values for $(\lecnn,\hmclecs)$. Equation~\eqref{eq:lec_posterior} for the sought posterior \pdf\ of the LECs thus becomes
\begin{equation}
\begin{split}
\label{eq:posterior}
\prob(\lecs | D, I) \propto&\ \prob(\aexpnn | \lecnn, \hmclecs, I) \\
&\times \prob(\lecnn | I) \times \prob(\hmclecs | \hmcdata, I).
\end{split}
\end{equation}

One could certainly argue, using previous knowledge of IB, that we are in the right to place a narrow prior for $\lecnn$ based on the marginal \pdf s for $\Ctnp$ and $\Ctpp$. However, to avoid building in such expectations on IB we selected a rather weak, and normally distributed, prior according to
\begin{equation}
  \prob(\lecnn | I) = \normal(0, \bar{\lec}_{nn}^2)
  \label{eq:lecnnprior}
\end{equation}
of width $\bar{\lec}_{nn} = 5$ $\Ctunit$.

We develop the likelihood in~\eqref{eq:posterior} by relating a theoretically computed value $\athnn$ of the \nn\ scattering length to the experimental result via a stochastic model
\begin{equation}
\label{eq:exp_and_th}
\aexpnn = \athnn + \delta \aexpnn + \delta \athnn.
\end{equation}
This relation introduces the experimental error $\delta \aexpnn$ and the \chieft{} error $\delta a\subth\supnn$, which we assume is dominated by the truncation of the \chieft{} series. Equation~\eqref{eq:exp_and_th} implies that other errors---such as numerical errors---are negligible. The assumption that $\delta \aexpnn$ and $\delta \athnn$ are independent and normally distributed random variables leads to a Gaussian likelihood for the scattering length,
\begin{equation}
  \prob(\aexpnn | \lecnn, \hmclecs, I) \propto \exp(-\frac{(\aexpnn - \athnn)^2}{2 (\sigma\subexp^2 + \sigma\subth^2)}),
  \label{eq:aexpnnL}
\end{equation}
where $\sigma\subexp^2$ and $\sigma\subth^2$ denote the variances of the experimental and theoretical errors, respectively. We use $\sigma\subexp = 0.4$ fm~\cite{Chen:2008zzj,Gardestig:2009ya,Gobel:2021pvw} as the experimental error.

To model the truncation error, we use the procedure from Refs.~\cite{Epelbaum:2014efa,Furnstahl:2015rha,Wesolowski:2018lzj}. We therefore assume that we can express the order-by-order predictions for $\athnn$ as the sum
\begin{equation}
\label{eq:eftsum}
  \athnn = a\subref \sum_{i=0}^k c_i Q^i
\end{equation}
where $k$ is the chiral order, i.e., $k=0$ is \lo\, and $k=2, 3, 4, \ldots$ corresponds to \nlo, \nnlo, \nnnlo, $a\subref$ is a dimensionful reference value for the scattering length, $c_i$ are dimensionless \chieft{} expansion coefficients, and the \chieft{} expansion parameter is assumed to be given by $Q = m_\pi / \Lambda_b$, which is reasonable for a quantity defined in the zero-momentum limit and analyzed in a pionful theory. To simplify notation, we omit explicit reference to the $k$ dependence of $\athnn$. We employ a breakdown scale $\Lambda_b=600$ MeV in accordance with the first analysis we performed in~\cite{Svensson:2021lzs}. Assuming that the expansion coefficients $c_i$ (between and within chiral orders) are independent and normally distributed yields the following form for the truncation error:
\begin{equation}
\prob(\delta \athnn | \cbarsq, Q, a\subref, I) = \normal(0, \sigma\subth^2)
\end{equation}
where $\cbarsq$ is the variance that characterizes the magnitude of the \chieft{} expansion coefficients for the scattering length and effective range. One can show~\cite{Furnstahl:2015rha,Wesolowski:2018lzj} that
\begin{equation}
\sigma\subth^2 = \cbarsq a\subref^2 \frac{Q^{2(k+1)}}{1-Q^2}.
\end{equation}

For the analysis of the scattering length, we place a (conjugate) inverse-gamma ($\invgamma$) prior $\cbarsq$, leading to a marginal posterior for $\cbarsq$ which also follows an $\invgamma$ distribution but with updated values for the parameters~\cite{Melendez:2019izc}. Marginalization of the variance leads to a Student's $t$ distribution~\cite{gosset1908} for the truncation error (see Appendix~\ref{app:t_dist}). Thus, we have that the $\invgamma$ prior for $\cbarsq$, with initial parameters $\IGaprior$ and $\IGbprior$,
\begin{equation}
\prob(\cbarsq | I) = \invgamma(\IGaprior, \IGbprior) \propto \frac{1}{\cbar^{2(\IGaprior+1)}} \exp(-\frac{\IGbprior}{\cbarsq}),
\end{equation}
is updated by a set of observed expansion coefficients $\vec{c}$ according to
\begin{equation}
\label{eq:cbar_posterior}
\prob(\cbarsq | \vecc, I) = \invgamma(\IGa, \IGb) \propto \frac{1}{\cbar^{2(\IGa+1)}} \exp(-\frac{\IGb}{\cbarsq})
\end{equation}
with
\begin{equation}
\begin{split}
  \label{eq:ig_post_hyper}
  \IGa &= \IGaprior + \frac{n_c}{2} \\
  \IGb &= \IGbprior + \frac{\veccsq}{2}
\end{split}
\end{equation}
where $n_c$ is the length of the vector $\vecc$. In practice, we are of course rather limited in the amount of data that we have available to infer $\lecnn$ and learn about the corresponding \chieft{} error for $\athnn$. To that end we here exploit the order-by-order convergence pattern of $\athpp$ and $\athnp$ to learn about the \chieft{} truncation error in $\athnn$. As will be further discussed in Sec.~\ref{sec:sampling}, we consider the \nnlo-\nnnlo\ shift of $a\subth\suppp$ to be an outlier and therefore omit this from the order-by-order data used to learn about the magnitude of the EFT truncation error. In detail, we have $n_c = 5$ expansion coefficients for informing the marginal posterior for $\cbarsq$; these are calculated (see, e.g., Ref.~\cite{Svensson:2021lzs}) from the \lo-\nlo, \nlo-\nnlo shifts of $a\subth\supnp$ and $a\subth\suppp$, and the \nnlo-\nnnlo\ shift of $a\subth\supnp$, all evaluated at the maximum a posteriori (MAP) locations for the \pdf\ $\prob(\hmclecs | \hmcdata, I)$; see Sec.~\ref{sec:sampling} for details. However, with our priors, we find that the inference does not change dramatically if we include the \nnlo-\nnnlo\ shift of $a\subth\suppp$. This is described further in Sec.~\ref{sec:results} and Appendix~\ref{app:ppd_all}.

\begin{figure}[t]
\centering
\includegraphics[width=1.0\columnwidth]{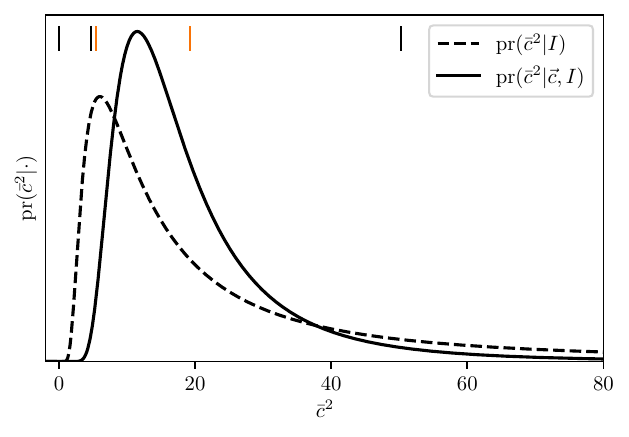}
\caption{The prior $\prob(\cbarsq |I)$ (dashed line) for $\cbarsq$ and the resulting marginal posterior $\prob(\cbarsq|\vecc,I)$ (solid line). The latter is obtained from allowing the order-by-order predictions of $a\subth\supnp$ and $a\subth\suppp$ to flow through the prior via the observed expansion coefficients $\vecc$. The individual (squared) elements of $\vecc$ are shown as vertical lines, with black (orange) representing \np\ (\pp) elements. The center dot in the $y$-axis label $\prob(\cbarsq|\cdot)$ acts as a placeholder for ``$I$'' and ``$\vecc, I$''.}
\label{fig:invgamma}
\end{figure}

We choose $(\IGaprior, \IGbprior) = (1.0, 12.0)$, which yields the prior and resulting marginal posterior shown in Fig.~\ref{fig:invgamma}. Our prior is predicated on our previous experience of these expansion coefficients, albeit for \NN\ scattering data: we see it as unlikely that $\cbarsq < 4$, but otherwise acknowledge our ignorance of the size of the truncation error. The mode of the prior, given by $\IGbprior / (\IGaprior + 1)$, is located at $\cbarsq = 6$. Exposure to data $\vecc$ shifts the bulk of the \pdf\ towards larger truncation errors, with the mode of the marginal posterior falling at $\cbarsq = 11.5$.

To account for the marginal \pdf\ on the \chieft{} truncation error, the posterior in~\eqref{eq:posterior} is expanded, viz.,
\begin{equation}
\prob(\lecs | \vecc, D, I) = \int \prob(\lecs, \cbarsq| \vecc, D, I) \, d \cbarsq,
\end{equation}
where we now explicitly indicate that the posterior is conditional on $\vecc$. In slightly more detail, the joint posterior of $\lecs$ and $\cbarsq$,
\begin{equation}
\begin{split}
\label{eq:joint_posterior}
\prob(\lecs, \cbarsq | \vecc, D, I) &\propto \prob(\aexpnn | \lecnn, \hmclecs, \cbarsq, I) \times \prob(\lecnn | I) \\
&\times \prob(\hmclecs | \hmcdata, I) \times \prob(\cbarsq | \vecc, I),
\end{split}
\end{equation}
conditioned on scattering data, scattering lengths, and order-by-order information, is the object of interest that we end up evaluating numerically.
Concluding this section we list possible extensions to our analysis that we leave for future work:
(1) incorporating a finite correlation length between the \chieft{} expansion coefficients as function of energy, (2) allowing for LEC variability, $c_i=c_i(\lecnn)$, (3) modifying the \chieft{} expansion parameter $Q$ to a slightly greater value $m_\pi/\unit[500]{MeV}$~\cite{Svensson:2021lzs,Wesolowski:2021cni} or, better yet, (4) account for the uncertainty in $Q$ by using an accompanying prior that ensures smooth matching to external (soft) momenta $>m_{\pi}$ in \chieft.

\subsection{Evaluating posteriors}
\label{sec:sampling}
In this section we expound on our sampling of the posterior for $\hmclecs$ using HMC and how we combine this posterior with $\lecnn$ to produce a joint posterior~\eqref{eq:joint_posterior} for all LECs at a given order. For clarity, we now make explicit that the inference of $\hmclecs$ is conditional on a fixed  variance \hmccbarsq{} of the \chieft{} expansion coefficients. A detailed account of how to efficiently sample \hmcpr{} using HMC was given in Ref.~\cite{Svensson:2021lzs}. Here, we will mainly remark on new developments and results. Our procedure for extracting the joint posterior can be considered a two-step process:
\begin{enumerate}
\item Sample the posterior \hmcpr{} using HMC.
\item Numerically evaluate the joint posterior \fullpr{} in~\eqref{eq:joint_posterior} using importance sampling.
\end{enumerate}

\subsubsection{Step 1: Sampling \hmcpr{} using HMC \label{sec:sampling_using_hmc}}
We follow the steps laid out in Ref.~\cite{Svensson:2021lzs}, with two modifications. First, we consistently use all available chiral orders, i.e., up to \nnnlo, to learn about the variance of the truncation error. Second, we employ a new method for rapidly tuning the parameters of the HMC algorithm such that we achieve sufficiently high sampling efficiency to perform reliable sampling at \nnnlo. This yields an \nnnlo\ posterior that passes all imposed MCMC convergence tests.

Let us first focus on how we learn about the truncation error. In our previous work~\cite{Svensson:2021lzs} we limited ourselves to only use information about the \chieft{} convergence pattern up to the order at which we were sampling to estimate the variance $\cbarsq$ of the truncation error, and we included the zeroth-order coefficients $\vec{c}_0$ in the estimation. We then concluded that this procedure typically leads to an underestimation of the size of the truncation error. Here, we follow our own advice and infer the variance $\cbarsq$ from all orders available to us, and exclude the rather uninformative and biasing zeroth-order coefficients. We use the same grid of observables, laboratory energies, and scattering angles, as in our previous work and arrive at $\hmccbar = 4.1$. This value is somewhat greater than the nominally expected natural scale of $\cbar \approx 1$, yet nothing too alarming and certainly in line with what we have observed before. As discussed in Sec.~\ref{sec:setting_up_the_inference}, this is also the basis for us shifting the prior for the \chieft{} truncation error for the effective range towards slightly greater values.

The importance of tuning the HMC parameters, along with various strategies for achieving efficient sampling, is detailed in Ref.~\cite{Svensson:2021lzs}. In particular, we stress the importance of a suitable so-called mass matrix: a matrix that accounts for differences in scale between parameters in the sampled \pdf. Previously we extracted a performant mass matrix by executing a short preliminary sampling and inverting the covariance matrix of the samples. Here, we instead optimize the LECs with respect to the posterior, i.e., we solve for
\begin{equation}
  \lecs_{\star} = \argmax_{\hmclecs} \,\, \hmcpr,
  \label{eq:map}
\end{equation}
and estimate the parameter covariance matrix at that optimum. We use the BFGS optimization algorithm~\cite{broyden70,fletcher70,goldfarb70,shanno70} which is a so-called quasi-Newton method that relies on first-order gradients to update approximations to the Hessian employed in the Newton algorithm.  We employ automatic differentiation (AD)~\cite{Griewank:2003,charpentier09,Carlsson:2015vda} to obtain the first-order gradients. In the end, we find that the approximate Hessian at the optimum, as found by the BFGS algorithm, makes for a suitable mass matrix in our application. In our case we have access to higher-order gradients via AD and could, in principle, obtain $\lecs_{\star}$ using any expedient optimization method and compute an exact-to-machine-precision Hessian. However, we find that BFGS is sufficient for tuning the HMC sampler. Bypassing the need for the $\approx 20\,000$ function evaluations required during a preliminary HMC sampling is certainly a great improvement, as the BFGS algorithm typically terminates after at most a few hundred function evaluations. Better yet, the optimization-based method is in our experience more reliable and ultimately yields slightly higher performance in the subsequent sampling of the posterior.

We gather $N\approx 10\,000$ samples in each of the ten chains we run at each chiral order. We find integrated autocorrelation times $\tau$ ($\tau_\textnormal{\nlo} = 0.71$, $\tau_\textnormal{\nnlo} = 0.93$, and $\tau_\textnormal{\nnnlo} = 3.33$) that reach a plateau within very few samples, thus indicating converged HMC chains. The $\tau$ value should not be too great since it is inversely proportional to the effective sample size that measures the number of independent samples in the HMC chain. We note that the \nlo\ and \nnlo\ autocorrelation times are similar to our previous results~\cite{Svensson:2021lzs}, achieved with the more cumbersome method of preliminary samplings. We should reiterate the concerns that always surround the topic of convergence in MCMC. In real-world applications it is not possible for us to declare a chain ``converged''. This is because we can never explore the entire parameter space in finite time, and we can miss non-negligible (or even dominating) probability regions. All we can do is to probe the MCMC chains for signs of non-convergence. This issue becomes progressively more pressing as the dimensionality and domain of the parameter space grows, and is especially concerning at \nnnlo\ in our case. With that said, we have searched the parameter domains to the best of our abilities by initializing the BFGS optimization, and the HMC algorithm, at multiple locations and we did not find any signs of multimodality and non-convergence.

In Table~\ref{tab:ere_alpha_star} we predict the scattering lengths and effective ranges for the \np\ and \pp\ isospin channels in the $^{1}S_0$ partial wave at the MAP points~\eqref{eq:map} of the LEC posteriors at different orders. Overall, the predictions are reasonably close to the experimental, or rather the empirical, values~\cite{Machleidt:2011zz}. This builds confidence in our model inference. These point estimates give a first indication of our model's performance in the low-energy limit. Overall, we observe convergence towards empirical results as we move to higher chiral orders. A notable exception is the \nnnlo\ prediction of the scattering length in the \pp\ isospin channel. It differs from the empirical result by 1.9 standard deviations, compared to just 1.2 standard deviations for the \nnlo\ prediction. Furthermore, the difference is in the opposite direction. This is possibly caused by a lack of informative data during the inference of the LECs.

\begin{table}
\caption{MAP predictions of \np\ and \pp\ scattering lengths ($a$) and effective ranges ($r$) in units of fm. Empirical results are from Ref.~\cite{Machleidt:2011zz}.}
\label{tab:ere_alpha_star}
\begin{tabular}{ c | c c c c }
Order & $a_{np}$ & $a_{pp}$  & $r_{np}$  & $r_{pp}$   \\
\hline
\lo   & $-24.428$ & $-24.056$ & $1.766$ & $1.768$ \\
\nlo  & $-21.625$ & $-18.476$ & $2.497$ & $2.585$ \\
\nnlo & $-23.732$ & $-17.797$ & $2.681$ & $2.814$ \\
\nnnlo& $-23.733$ & $-16.544$ & $2.693$ & $2.867$ \\
\hline
Empirical & $-23.740(20)$ & $-17.3(4)$ & $2.77(5)$ & $2.85(4)$ \\
\end{tabular}
\end{table}

\subsubsection{Step 2: Evaluating the joint posterior}
\label{sec:sampling_full}
Equipped with an MCMC chain to represent \hmcpr{}, we proceed to evaluate the posterior \fullpr{} in Eq.~\eqref{eq:joint_posterior}. Due to the expected weakness of IB effects in our \chieft{} we can be fairly certain that the posterior probability mass of the latter \pdf{} will not be very far from the one of the former. We take advantage of this expectation and proceed by using the principles of sampling/importance sampling~\cite{smith92} where we first define a sampling distribution
\begin{equation}
\begin{split}
  g(\lecs, \cbarsq) \equiv&\ \hmcpr \\
&\times \prob(\cbarsq | \vecc, I) \times \pi(\lecnn),
  \label{eq:g}
\end{split}
\end{equation}
with $\pi(\lecnn)$ a simple, bounded uniform distribution. Since the likelihood for the single \nn{} scattering length poses no significant computational challenge, for each chiral order we first explore the $\lecnn{}$ space to identify an interval of values encompassing the bulk of the
probability mass. These intervals are (in units of $10^4\,\textnormal{GeV}^{-2}$): $\Ctnn \in [-0.1495,-0.1445]$ at \nlo, $\Ctnn \in [-0.1535,-0.1510]$ at \nnlo, and $\Ctnn \in [-0.132,-0.125]$ at \nnnlo. We then sample the full posterior~\eqref{eq:joint_posterior} using the following procedure:
\begin{enumerate}
\item Pick a sample $(\lecs{}_i,\cbarsq_i)$ from the sampling distribution~\eqref{eq:g}. In practice, we go through the entire MCMC chain of $\hmclecs$ sequentially since we know~\cite{Svensson:2021lzs} that these are independent and random samples from \hmcpr. For each \hmclecs{} sample we draw a sample of $\cbarsq$ from its marginal posterior $\prob(\cbarsq | \vecc, I)$, see Eq.~\eqref{eq:cbar_posterior}, and a sample of $\lecnn$ from $\pi(\lecnn)$.
\item Evaluate the ratio $\omega_i = \fullpr{i} /g (\lecs{}_i,\cbarsq_i)$. Due to the form of our sampling distribution~\eqref{eq:g} and the factorization of the joint posterior~\eqref{eq:joint_posterior}, this ratio is simply the product of the Gaussian likelihood for the scattering length~\eqref{eq:aexpnnL} and the prior for $\lecnn$~\eqref{eq:lecnnprior}.
\end{enumerate}
Once we have the lists of samples $(\lecs_i, \cbarsq_i)$ we compute the normalized (importance) weights $q_i = \omega_i / \sum_j \omega_j$. This provides us with a weighted chain of $(\lecs, \cbarsq)$ values distributed as the target posterior \fullpr{}~\eqref{eq:joint_posterior}. The method described here is simple to apply but may not work well in all cases, e.g., if the marginal posterior of $\lecnn$ is relatively unknown and/or defined for a high-dimensional parameter domain. We have validated our results obtained using importance sampling by approximating $\hmcpr$ with a multivariate normal distribution and directly sampling the full distribution~\eqref{eq:joint_posterior} using HMC. Multivariate normal approximations to all posteriors are provided in the Supplemental Material~\cite{suppl}.

\begin{table*}
\caption{MAP predictions of few-nucleon ground-state energies $E$ (in MeV) and point-proton radii $R$ (in fm) for nuclei with mass number $A=2,3,4$ obtained using the Jacobi no-core shell-model~\cite{Navratil:1999pw} in a harmonic oscillator basis with frequency 22 MeV$/\hbar$ in model spaces with 251, 41, and 21 oscillator shells, respectively. Note that these predictions are not including a three-nucleon (\NNN) interaction. For $^2$H we also list the $D$-state probability ($P_D$) in \% and the electric quadrupole moment (Q) in units of $e$b. The experimental and empirical values are the same as in Ref.~\cite{Carlsson:2015vda}.}
\label{tab:ncsm_alpha_star}
\begin{tabular}{ c | c c c c c c c c c c}
Chiral order& $E(^2$H) & $R(^2$H) & $P_D(^2$H)& $Q(^2$H) & $E(^3$H)& $R(^3$H)  & $E(^3$He)  & $R(^3$He)  & $E(^4$He)  & $R(^4$He) \\
\hline
\lo   & $-0.855$   & $2.854$    & $5.347$&  $0.411$       & $-7.214$ & $1.510$     & $-6.327$ & $1.598$ & $-28.781$&  $1.201$  \\
\nlo  & $-1.810$   & $2.105$    & $3.052$&  $0.282$       & $-8.311$ & $1.544$     & $-7.559$ & $1.700$ & $-30.233$&  $1.342$  \\
\nnlo\ (\NN{} only) & $-2.165$   & $1.982$    & $3.212$&  $0.267$       & $-8.570$ & $1.565$     & $-7.791$ & $1.745$ & $-29.873$&  $1.376$  \\
\nnnlo\ (\NN{} only) & $-2.268$   & $1.978$    & $3.476$&  $0.278$       & $-7.560$ & $1.715$     & $-6.829$ & $1.918$ & $-24.272$&  $1.570$  \\
\hline
Experiment and empirical & $-2.225$ & $1.976(1)$ & --      & $0.270(11)$ & $-8.428$ & $1.587(41)$ & $-7.718$ & $1.766(5)$& $-28.30$ & $1.455(6)$ \\
\end{tabular}
\end{table*}

We report bivariate and univariate marginals of the joint LEC posteriors in Appendix~\ref{app:lec_posteriors}. In Table~\ref{tab:ncsm_alpha_star} we make point-estimate predictions for selected few-nucleon ground-state observables using the MAP point of the joint posteriors. This is a first check of the model inference and the results are reasonable. We note that the \nnnlo\ predictions deviate more from experiment than \nnlo. However, we do not incorporate \NNN{} interactions in this analysis and for that reason we also refrain from estimating \chieft{} truncation errors for the results presented in Table~\ref{tab:ncsm_alpha_star}. On the other hand, we perform a full analysis of the \NN{} effective range expansion in Sec.~\ref{sec:results}.

Apart from the inclusion of $\Ctnn$ and $\cbar$, the \nlo\ and \nnlo\ posteriors are overall similar to their counterparts in Ref.~\cite{Svensson:2021lzs}. The differences that do occur arise prior to the inclusion of $\Ctnn$ as a result of different characterizations of the \chieft{} truncation error. The correlation patterns are nearly identical, and the majority of the marginal posteriors overlap at the 68\% level. However, some parameters in higher partial waves have somewhat shifted values. For example, in Ref.~\cite{Svensson:2021lzs} we reported $C_{3P2} = -0.162(1) \Cunit$ at \nlo, while Fig.~\ref{fig:posterior_nlo} reveals that $C_{3P2} = -0.175(2) \Cunit$.

The posteriors presented here are confined to rather small volumes in parameter space and, as expected, the marginal posteriors for the $\Ct_{1S0}$ LEC acting in the different isospin channels are similar to each other. The marginal posterior for $\Ctnn$ is roughly twice as wide as those for $\Ctnp$ and $\Ctpp$. This can largely be attributed to the vastly more abundant \np\ and \pp\ scattering data. The posteriors pick up correlations between $\Ctnn$ and $\Ctnp$, $\Ctpp$. At \nnnlo, $\Ctnn$ shows stronger correlations with other LECs than at \nlo\ and \nnlo. The fourth-order contact LECs, $D$, are sensitive to high momentum observables and therefore less constrained by the data. We find that several of the $D$ LECs return the prior if we do not condition the posterior on data in the $\Tlab \in [40.5, 290]$ MeV region. Furthermore, the posteriors indicate that some $D$ LECs are notably unnatural; for instance, we have $D_{1S0} = -29.1(3) \Dunit$, which is of particular interest in this case as it acts in the same partial wave as---and is strongly correlated with---our explicitly isospin breaking LECs. The inference of this LEC, then, is influential on our predictions of $^1S_0$ scattering lengths and may explain the stark difference between the \nnnlo\ predictions for $a\subnp$ and $a\subpp$ in Table~\ref{tab:ere_alpha_star}. There are precedents for large values of $D_{1S0}$; see, e.g., Ref.~\cite{Entem:2003ft}. Unlike earlier works, we also find rather unnaturally sized MAP values for $D_{3S1}=-32.2(3) \Dunit$ and $D_{3P1}=-20.4(1) \Dunit$, where the uncertainties denote 68\% credible intervals. At \nnnlo, the various $\Ct$ and $C$ parameters are akin to their values inferred at \nnlo, with the exception of $C_{3P1}$ which is both several times larger and has the opposite sign, i.e., $C_{3P1}=-0.956(5) \Cunit$ at \nnlo\ and $C_{3P1}=6.13(1) \Cunit$ at \nnnlo.

We have also investigated how the posteriors are affected if we do not exclude the unexpectedly large \nnlo-\nnnlo\ shift in $a\subth\suppp$ from the estimation of the truncation error (see Sec.~\ref{sec:setting_up_the_inference}). We find that the width of the marginal \Ctnn\ posterior roughly doubles at \nlo, widens by a small but noticeable amount at \nnlo, and is virtually unaffected at \nnnlo. From this we draw the conclusion that the theoretical uncertainty dominates at \nlo, the theoretical and experimental uncertainties are roughly equal at \nnlo, and the experimental uncertainty dominates at \nnnlo.

\begin{figure}[tb]
\centering
\includegraphics[width=1.0\columnwidth]{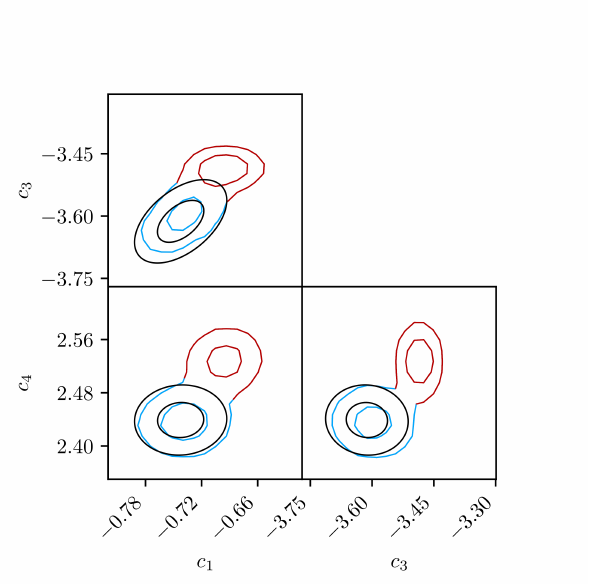}
\caption{Prior and posterior \pdf s for the \piN{} LECs $c_1,c_3,c_4$, in units of GeV$^{-1}$. The posteriors were obtained using HMC. The inner (outer) ellipses enclose approximately 39\% (86\%) of the probability mass. Black lines indicate the prior as provided by the Roy-Steiner analysis~\cite{Siemens:2016jwj}. Blue lines indicate the posterior conditioned on low-energy scattering data only, i.e., with $\Tlab$ up to 40.5 MeV. Red lines indicate the posterior conditioned on scattering data up to $\Tlab = 290$ MeV.}
\label{fig:piN_contours}
\end{figure}

We find nonoverlapping priors and posteriors for the \piN\ LECs at both \nnlo\ and \nnnlo. This tension was previously seen at \nnlo\ in Ref.~\cite{Svensson:2021lzs}. It turns out that if the posteriors are conditioned on only a low-energy data set, $\Tlab \in [0,40.5]$ MeV, we return our priors from a Roy-Steiner analysis. This is summarized for the \nnlo\ case in Fig.~\ref{fig:piN_contours}. Although there are significant deviations between the \piN\ prior and posterior when we condition on high-energy data, the discrepancies are rather small on a naturalness scale. As such, the statistical model we have set up to relate low-energy data and \chieft\ at \nnlo\ appears to preserve the long-range physics rather well. Moreover, the credible intervals of the \piN\ posterior cannot be straightforwardly compared with the variance of the prior determined in a Roy-Steiner analysis~\cite{Hoferichter:2015hva}, for which it is difficult to estimate the truncation uncertainty. At \nnnlo\ we similarly find that conditioning on the energy-truncated data set returns the prior for the \piN\ LECs. The prior and posterior at \nnnlo\ (conditioned on the full data set) are shown side by side in Appendix~\ref{app:lec_posteriors}. It is peculiar that the marginal and univariate posterior for the \piN{} LEC $c_4$ sits on top of the prior.

\section{Posterior predictive distributions of $\boldsymbol{^1S_0}$ isospin breaking }
\label{sec:results}
\begin{figure*}[ht!]
\centering
\includegraphics[width=1.0\linewidth]{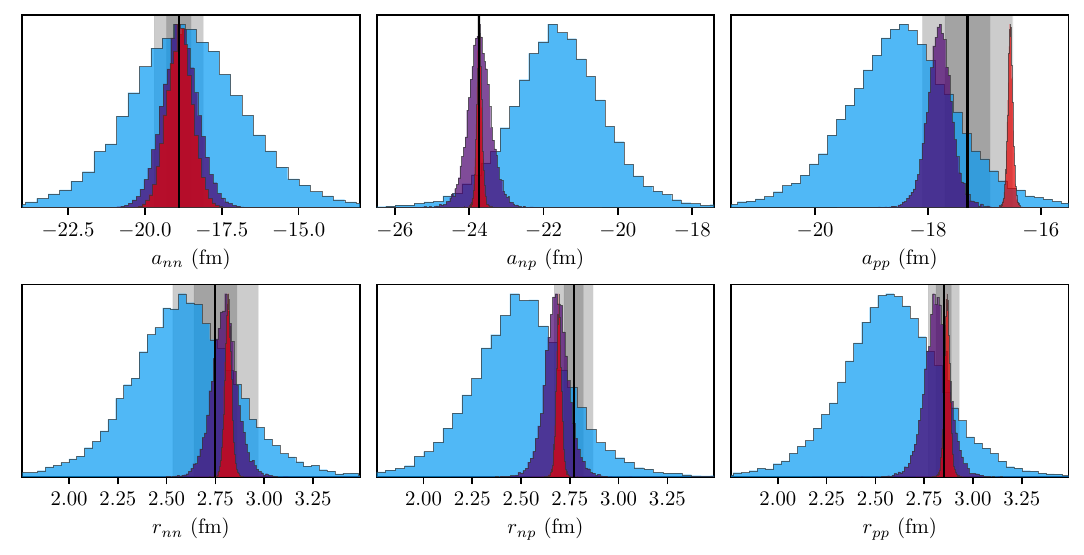}
\caption{\ppd s of scattering lengths and effective ranges at \nlo\ (blue), \nnlo\ (purple), and \nnnlo\ (red). Empirical results are shown as black lines, with corresponding $1\sigma$ $(2\sigma)$ uncertainties as a dark (light) gray area. The empirical results are without electromagnetic effects and gathered from Ref.~\cite{Machleidt:2011zz}, except for $a\subnn$ which we take from Ref.~\cite{Gardestig:2009ya}. Note that the histograms have been scaled such that they have the same peak height.}
\label{fig:ppd_panel}
\end{figure*}
Equipped with samples from the joint posterior \pdf s of the \NN{} LECs up to \nnnlo\ we proceed to sampling the posterior predictive distributions (\ppd s) for the $S$-wave scattering length $a$ and effective range $r$ defined by the effective range expansion (ERE)
\begin{equation}
  p \cot \delta(p) = -\frac{1}{a} + \frac{1}{2}rp^2 + \mathcal{O}(p^4).
  \label{eq:ERE}
\end{equation}
Here, $p$ is the relative momentum and $\delta(p)$ is the scattering phase shift. The \ppd\ is the \pdf\ of unobserved values for $a$ and $r$ conditioned on \np\ and \pp\ scattering data as well as the empirical \nn{} scattering length.

We quantify the \ppd{} in all three isospin channels $t_z=$\pp,\np,\nn{} of the $^{1}S_0$ partial wave. The full \ppd s are given by the finite set of samples
\begin{equation}
  \left\{a^{t_z}_\text{th}(\lecs) + \delta a_\text{th}\right\}
  \label{eq:ppd_ath}
\end{equation}
and
\begin{equation}
  \left\{r^{t_z}_\text{th}(\lecs) + \delta r_\text{th}\right\},
  \label{eq:ppd_rth}
\end{equation}
with $(\lecs, \cbarsq) \sim \fullpr{}$. Furthermore, the conjugacy of the $\invgamma$ prior for the \chieft{} truncation error variance enables a pointwise and closed-form evaluation of the associated \chieft{} truncation errors $\delta a_\text{th}$ and $\delta r_\text{th}$; see Ref.~\cite{Melendez:2019izc} and Appendix~\ref{app:t_dist}. As a result of our sampling procedure (see Sec.~\ref{sec:sampling}) the sets in Eqs.~\eqref{eq:ppd_ath} and \eqref{eq:ppd_rth} are composed of approximately $10^5$ weighted samples. In the following we present results based on an inference of $\cbar$ where we excluded the \nnlo-\nnnlo\ shift of the $a\subth\suppp$ (see Table~\ref{tab:ere_alpha_star}). It turns out that this does not impact any predictions beyond \nlo; see Appendix~\ref{app:ppd_all} for comparison.

The results at \nlo, \nnlo, and \nnnlo\ are shown in Fig.~\ref{fig:ppd_panel}. Clearly, our model is not fine-tuned to reproduce any of the ERE parameters in Fig.~\ref{fig:ppd_panel} except $a\subnn$, which therefore serves as another model check, but all \ppd s agree with the empirical results within their uncertainties and the different distributions at different orders overlap reasonably with each other. The exception here is the \nnnlo\ prediction of $a\subpp$, that we also omitted from the estimation of the \chieft{} truncation error. In contrast, for $a\subnp$, both \nnlo\ and \nnnlo\ agree perfectly with the rather precise empirical value. For the empirical values of the \pp{} ERE parameters, the error bands emerge predominantly from the model dependence in the analysis of scattering data and removal of electromagnetic effects. The \ppd s for the effective ranges show a great deal of congruity. This implies that the \NN\ contact $C_{^1S_0}$ that enters at \nlo\ with a quadratic momentum dependence is sensibly inferred. This contact is missing at \lo\ and the predictions of effective ranges at that order are thus rather poor, as seen in Table~\ref{tab:ere_alpha_star}. Overall, one must go beyond \nnlo\ to achieve predictions that are more precise than the corresponding empirical uncertainties.

\begin{figure*}[ht!]
\centering
\includegraphics[width=1.0\linewidth]{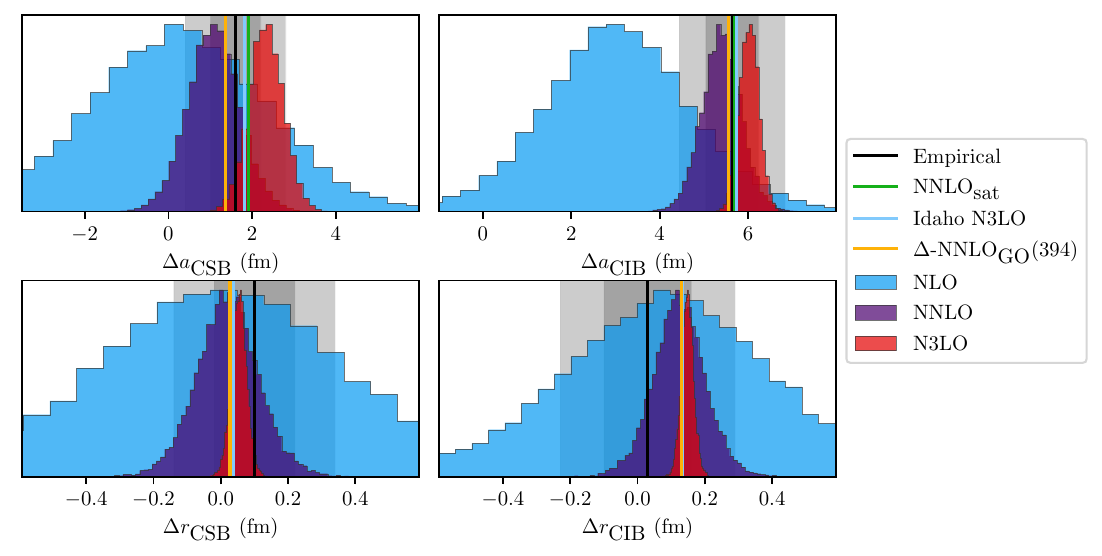}
\caption{CSB and CIB in the $^1S_0$ scattering lengths and effective ranges; see~\eqref{eq:csb_cib}. Colors and shadings as in Fig.~\ref{fig:ppd_panel}. Empirical results ($a\subnn$ from Ref.~\cite{Gardestig:2009ya}; all others from Ref.~\cite{Machleidt:2011zz}) are shown in black with gray error bands. Calculations using three widely used potentials are also shown: \nnlo$_\textnormal{sat}$~\cite{Ekstrom:2015rta}, Idaho-\nnnlo~\cite{Entem:2003ft}, and $\Delta$-\nnlo$_\textnormal{GO}(394)$~\cite{Jiang:2020the}. Note that the three latter potentials agree almost exactly regarding $\Delta r_\textnormal{CSB,CIB}$ and are thus difficult to visually distinguish. Also note that the histograms have been scaled such that they have the same peak height.}
\label{fig:csb_cib}
\end{figure*}

To quantify the strength of IB effects, we convert the \ppd s for the ERE parameters in the $^{1}S_0$ partial wave to standard measures of the CIB and charge symmetry breaking (CSB), where the latter amounts to a $\pi$ rotation around the $y$ axis in isospace, i.e.,
\begin{align}
  \label{eq:csb_cib}
  \begin{split}
    {}&\Delta a_\text{CIB} = \frac{1}{2}\left( a_\text{th}\suppp + a_\text{th}\supnn \right) - a_\text{th}\supnp ,\,\, \Delta a_\text{CSB} = a_\text{th}\suppp - a_\text{th}\supnn \\
    {}&\Delta r_\text{CIB} = \frac{1}{2}\left( r_\text{th}\suppp + r_\text{th}\supnn \right) - r_\text{th}\supnp ,\,\, \Delta r_\text{CSB} = r_\text{th}\suppp - r_\text{th}\supnn .
  \end{split}
\end{align}
The \ppd s for the CIB and CSB effects, including \chieft{} errors, are summarized in Fig.~\ref{fig:csb_cib}. We first note that it is only at \nnlo\ and beyond that we can detect, with confidence, the overall magnitudes of IB effects in the $^{1}S_0$ partial wave. Only CIB in the scattering length can be said to be different from zero with any confidence at \nlo. We also find that our \ppd s for CSB and CIB agree, within uncertainties, with existing empirical data and a range of point estimates using well-known chiral potentials at \nnlo\ and \nnnlo. These point estimates are all rather close to each other and fall within the empirical uncertainties at all orders. The CSB and CIB in the effective range is somewhat underestimated and overestimated, respectively, compared to the empirical values. The outlier \nnnlo\ result for $a\subpp$ of course propagates to the results for CIB and CSB and induces a comparatively large isospin breaking at this order. Yet, the results are in line with other chiral potentials.

\section{Summary and Outlook}
\label{sec:summary}
In this paper, we sampled high-dimensional posteriors up to \nnnlo\ in \chieft{} and studied the effects of IB in the \NN{} sector. We used Bayesian inference, conditioned on \np{} and \pp{} scattering data as well as the empirical value $\aexpnn$ for the $^{1}S_0$ scattering length in the \nn\ channel, to infer posterior \pdf s for the LECs at \lo, \nlo, \nnlo, and \nnnlo. We split the inference in two steps. First we employed HMC and conditioned the LEC posteriors on \np{} and \pp{} scattering data and accounted for uncorrelated \chieft{} truncation errors. A new approach to tuning the mass matrix, based on posterior optimization, enabled us to extract a converged 31-dimensional posterior also at \nnnlo. We find such advancements pivotal for enabling robust MCMC sampling of the \nnnlo\ posterior. In the next step of the inference, we included $\aexpnn$ and employed importance sampling to marginalize in the \nn\ contact LEC. For the variance of the truncation error in the ERE parameters in the $^{1}S_0$ partial wave we employed a conjugate $\invgamma$ prior. In the end, we find that the resulting LEC posteriors at \nlo\ and \nnlo\ match our previous~\cite{Svensson:2021lzs} posteriors in the \np\ and \pp\ isospin sectors and that the LEC $\Ctnn$ is roughly twice as broad but exhibits a correlation pattern similar to $\Ctnp$ and $\Ctpp$.

Next, we sampled the \ppd s for the ERE parameters and found that our results are consistent with existing point estimates using well-known chiral potentials, with the exception of $a\subpp$ at \nnnlo. This outlier might be traced back to insufficient information in the \NN{} scattering data set we conditioned the first part of the inference on. When accounting for \chieft{} truncation errors, we find that one must go to \nnlo\ to confidently detect IB effects.

One way to improve a Bayesian analysis of the strong interaction in the low-energy region of relevance to the ERE might be to mix pionless EFT~\cite{Bedaque:2002mn,Hammer:2019poc} and \chieft{}. Indeed, \chieft{} harbors an intrinsic uncertainty due to the low-energy scale set by $m_{\pi}$. Since the exact form of the \chieft{} expansion parameter $Q$ for scattering amplitudes is obscured by the non-perturbative resummation in the Lippmann-Schwinger equation its uncertainty $\prob(Q|I)$ should be accounted for. It could then be interesting to apply Bayesian mixture models to combine pionless EFT and \chieft\ predictions.

When we condition the inference on low-energy \NN{} scattering data with $\Tlab \in [0,40.5]$ MeV we return the Roy-Steiner priors for the \piN{} LECs at \nnlo\ and \nnnlo. This is in accordance with \chieft{} being a low-energy theory with long-ranged physics governed by the \piN{} interaction. When conditioning the inference on all \NN{} data up to the pion-production threshold at 290 MeV, the marginal \piN{} LEC posteriors are significantly shifted with respect to the priors. Still, the discrepancies are small on a naturalness scale.

\emph{Ab initio} studies~\cite{Ekstrom:2017koy,Jiang:2020the} indicate that the inclusion of the $\Delta$ isobar yields more realistic predictions for bulk properties of nuclei and nuclear matter. However, a Roy-Steiner prior for the \piN{} LECs with explicit $\Delta$ isobars is less precise~\cite{Siemens:2016jwj}. As such, it will also become more important to employ additional nuclear structure and reaction data in the inference and employ error models that account for correlations between the \chieft{} expansion coefficients. Although it rapidly becomes computationally challenging to evaluate likelihoods encompassing nuclear data for increasing mass numbers, the development of fast and accurate emulators~\cite{Konig:2019adq,Ekstrom:2019lss,Melendez:2021lyq,Zhang:2021jmi} appears to provide sufficient leverage.
In particular, the present work serves as an example how to sequentially incorporate multiple sets of low-energy nuclear structure and reaction data in a Bayesian framework to perform inference and quantify the theoretical precision.

\begin{acknowledgments}
This work was supported by the European Research Council (ERC) under the European Unions Horizon 2020 research and innovation program (Grant Agreement No. 758027), the Swedish Research Council (Grants No. 2017-04234 and No. 2021-04507). The computations were enabled by resources provided by the Swedish National Infrastructure for Computing (SNIC) partially funded by the Swedish Research Council through Grant Agreement No. 2018-05973
\end{acknowledgments}

\bibliography{main.bib}

\appendix

\section{The appearance of the Student's $t$ distribution}
\label{app:t_dist}
We assume that the $n_c$ expansion coefficients $\vecc = (c_1, \ldots, c_{n_c})$ in~\eqref{eq:eftsum} are independent and identically distributed, and drawn from a normal distribution with variance $\cbarsq$:
\begin{equation}
\label{eq:app_cdist}
\prob(c_i | \cbarsq) = \normal(0, \cbarsq).
\end{equation}
We then place an inverse-gamma ($\invgamma$) prior on $\cbarsq$, and consequently get an ($\invgamma$) posterior,~\eqref{eq:cbar_posterior}, as well due to conjugacy with respect to the normal distribution:
\begin{equation}
\prob(\cbarsq | \vecc) = \invgamma(\IGa, \IGb),
\end{equation}
where the parameters $\IGa$ (shape) and $\IGb$ (scale) have been updated, via~\eqref{eq:ig_post_hyper}, by the exposure to $\vecc$. We will now show that these assumptions result in a \ppd\ for an unseen coefficient $\cnew$ given by a Student's $t$ distribution, i.e.,
\begin{equation}
\prob(\cnew | \vecc) = t_\nu(0, \tau^2),
\end{equation}
where $\nu$ and $\tau$ are the degrees of freedom and scale, respectively, of the Student's $t$ distribution using a method of derivation that can be applied in many other situations where conjugate priors are employed.

In general, the \ppd\ $\prob(\cnew | \vecc)$ for a new datum $\cnew$ given a parameter (or parameters) $\cbarsq$ and observed data $\vecc$ can be expressed as
\begin{equation}
\prob(\cnew | \vecc) = \frac{\prob(\cnew | \cbarsq, \vecc) \times \prob(\cbarsq | \vecc)}{\prob(\cbarsq | \cnew, \vecc)},
\end{equation}
which is easily verified using the product rule of probabilities. We know that $\prob(\cbarsq | \vecc) = \invgamma(\IGa, \IGb)$, and $\prob(\cbarsq | \cnew, \vecc)$ is obviously also an inverse-gamma distribution with further updated parameters $(\IGanew, \IGbnew)$
\begin{equation}
  \begin{split}
    \label{eq:twice_update}
  \IGanew &= \IGa + \frac{1}{2} = \IGa_0 + \frac{n_c}{2} + \frac{1}{2}, \\
  \IGbnew &= \IGb + \frac{\cnew^2}{2} = \IGb_0 + \frac{\veccsq}{2} + \frac{\cnew^2}{2}.
\end{split}
\end{equation}
Since $\cbarsq$ is given in $\prob(\cnew | \cbarsq, \vecc)$, this is just a normal distribution as in~\eqref{eq:app_cdist}. Hence
\begin{equation}
\prob(\cnew | \vecc) = \frac{\normal(0,\cbarsq) \times \invgamma(\IGa, \IGb)}{\invgamma(\IGanew, \IGbnew)}
\end{equation}
and we obtain for the \ppd
\begin{align}
  \begin{split}
  \prob(\cnew | \vecc) = {}& \frac{1}{\sqrt{2\pi\cbarsq}} \exp \left(-\frac{\cnew^2}{2\cbarsq}\right) \\
  {}& \times \frac{\frac{\IGb^\IGa}{\Gamma(\IGa)} \times \cbar^{2(-\IGa-1)} \exp \left(-\frac{\IGb}{\cbarsq}\right)}{\frac{(\IGbnew)^{\IGanew}}{\Gamma(\IGanew)} \times \cbar^{2(-\IGanew-1)} \exp \left(-\frac{\IGbnew}{\cbarsq}\right)}.
  \end{split}
\end{align}
This is can be written as
\begin{equation}
\label{eq:app_part_1}
\prob(\cnew | \vecc) = \frac{1}{\sqrt{2\pi}} \frac{\IGb^\IGa \Gamma(\IGa_+)}{\IGb_+^{\IGa_+}\Gamma(\IGa)}.
\end{equation}
Using~\eqref{eq:twice_update} we can rewrite this further as
\begin{equation}
\label{eq:app_part_4}
\prob(\cnew | \vecc) = \frac{1}{\sqrt{2\pi\IGb}} \frac{\Gamma(\IGa + \frac{1}{2})}{\Gamma \left(\IGa \right)} \times \left(1+\frac{\cnew^2}{2\IGb}\right)^{-\IGa-\frac{1}{2}}.
\end{equation}
This is the \pdf\ for a Student's $t$ distribution with $\nu = 2\IGa$ degrees of freedom and scale $\tau = \sqrt{\IGb/\IGa}$, i.e.,
\begin{equation}
\prob(\cnew | \vecc) = t_\nu \left(0, \tau^2\right) = t_{2\IGa}\left(0, \frac{\IGb}{\IGa}\right).
\end{equation}
One can analogously show that the truncation error is given by
\begin{equation}
\prob(\delta a\subth | \vec{a}) = t_{2\IGa} \left(0, a\subref^2 \frac{Q^{2(k+1)}}{1-Q^2}\frac{\IGb}{\IGa}\right)
\end{equation}
where $\vec{a}$ is a set of observations of $a$ that can be transformed into expansion coefficients $\vecc$, and $\IGa$ and $\IGb$ are given by~\eqref{eq:ig_post_hyper}.

\section{LEC posteriors \label{app:lec_posteriors}}
\begin{figure*}[h]
\centering
\includegraphics[width=1.0\linewidth]{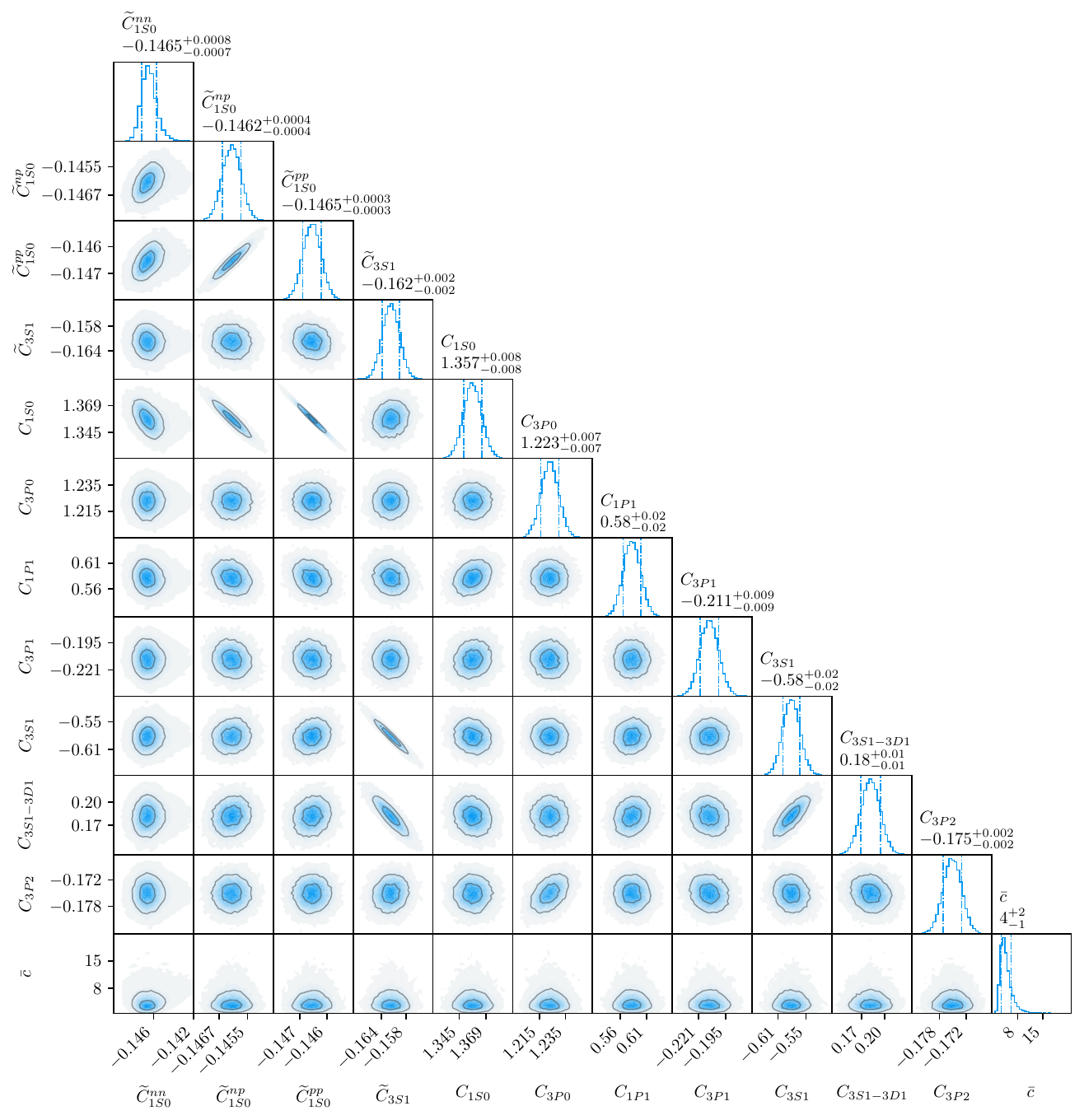}
\caption{The joint posterior $\prob(\lecs, \cbarsq | \nlo, \vecc, D)$ conditioned on \NN\ scattering data with $\Tlab \in [0,290]$ MeV, the $^{1}S_0$ \nn\ scattering length, and a conjugate prior for the \chieft{} truncation error of this quantity. The $\Ct$ LEC values are given in units of $10^4\,\textnormal{GeV}^{-2}$, and $C$ in units of $10^4\,\text{GeV}^{-4}$.}
\label{fig:posterior_nlo}
\end{figure*}

\begin{figure*}[h]
\centering
\includegraphics[width=1.0\linewidth]{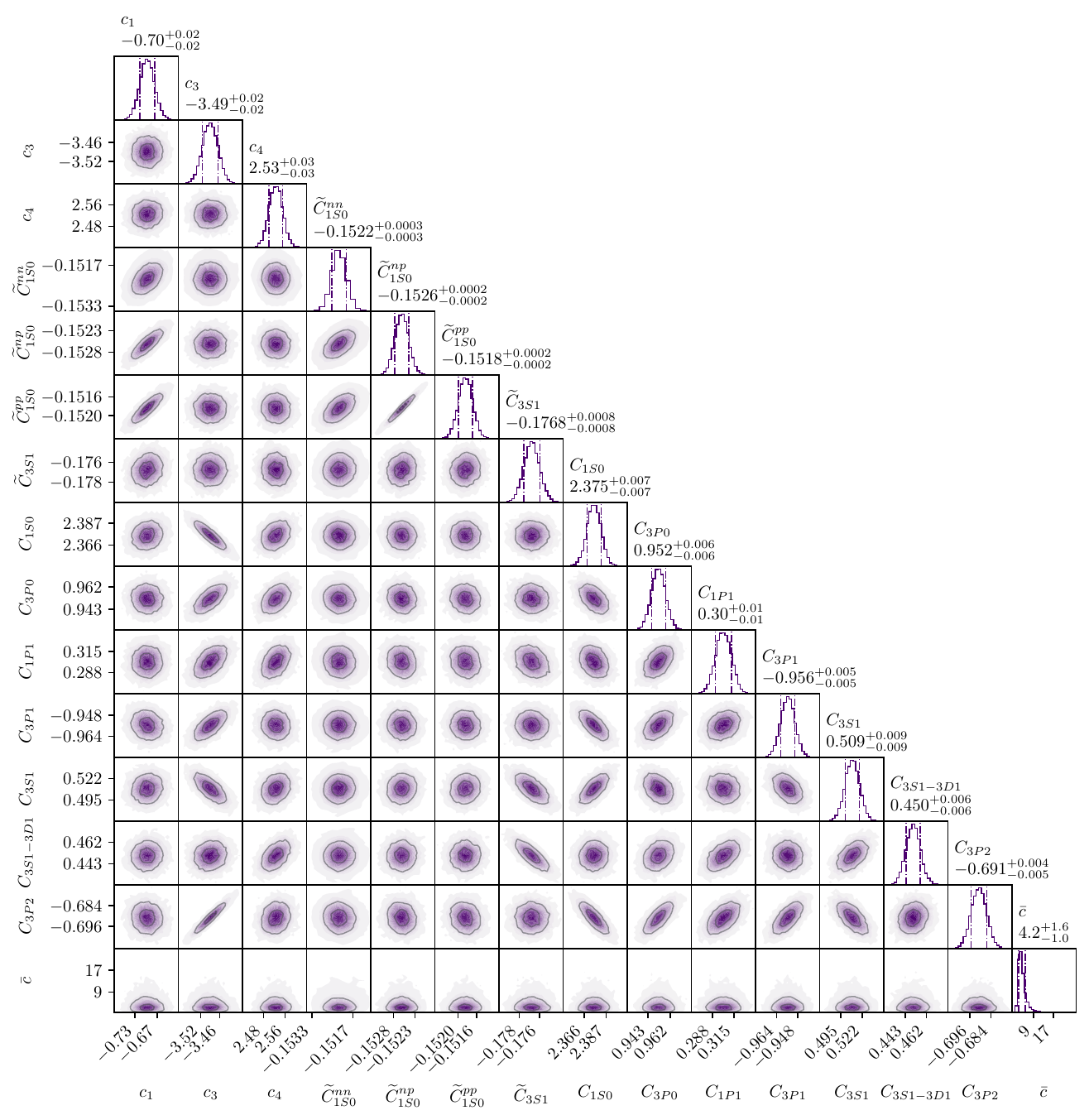}
\caption{The joint posterior $\prob(\lecs, \cbarsq | \nnlo, \vecc, D)$ conditioned on \NN\ scattering data with $\Tlab \in [0,290]$ MeV, the $^{1}S_0$ \nn\ scattering length, and a conjugate prior for the \chieft{} truncation error of this quantity. The $\Ct$ LEC values are given in units of $10^4\,\textnormal{GeV}^{-2}$, $C$ in units of $10^4\,\text{GeV}^{-4}$, and $c$ in units of GeV$^{-1}$.}
\label{fig:posterior_nnlo}
\end{figure*}

\begin{figure*}[h]
\centering
\includegraphics[width=1.0\linewidth]{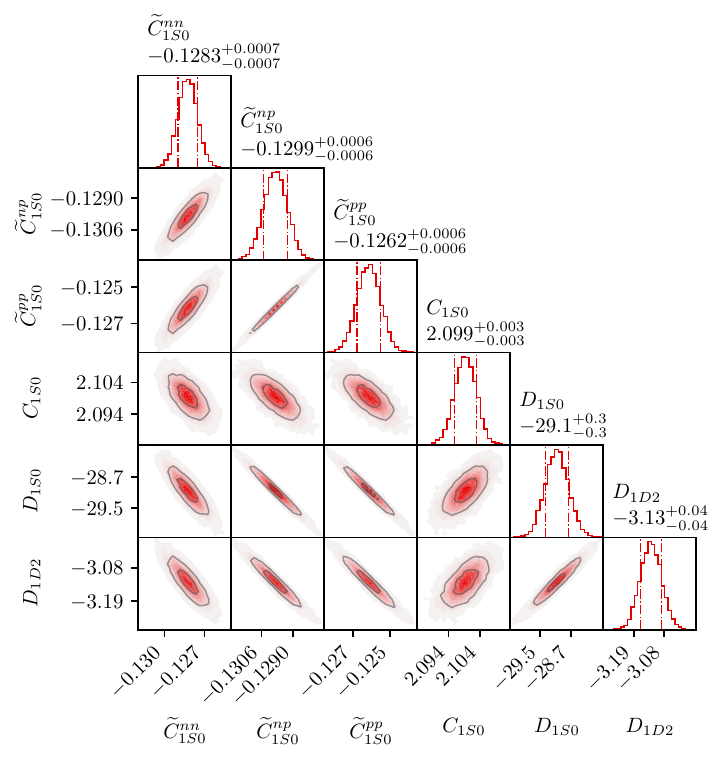}
\caption{A subset of the joint posterior $\prob(\lecs, \cbarsq | \nnnlo, \vecc, D)$ conditioned on \NN\ scattering data with $\Tlab \in [0,290]$ MeV, the $^{1}S_0$ \nn\ scattering length, and a conjugate prior for the \chieft{} truncation error of this quantity. The $\Ct$ LEC values are given in units of $10^4\,\textnormal{GeV}^{-2}$, $C$ in units of $10^4\,\text{GeV}^{-4}$, and $D$ in units of $10^4\,\text{GeV}^{-6}$.}
\label{fig:posterior_n3lo_extract}
\end{figure*}

\begin{figure*}[h]
\centering
\includegraphics[width=1.0\linewidth]{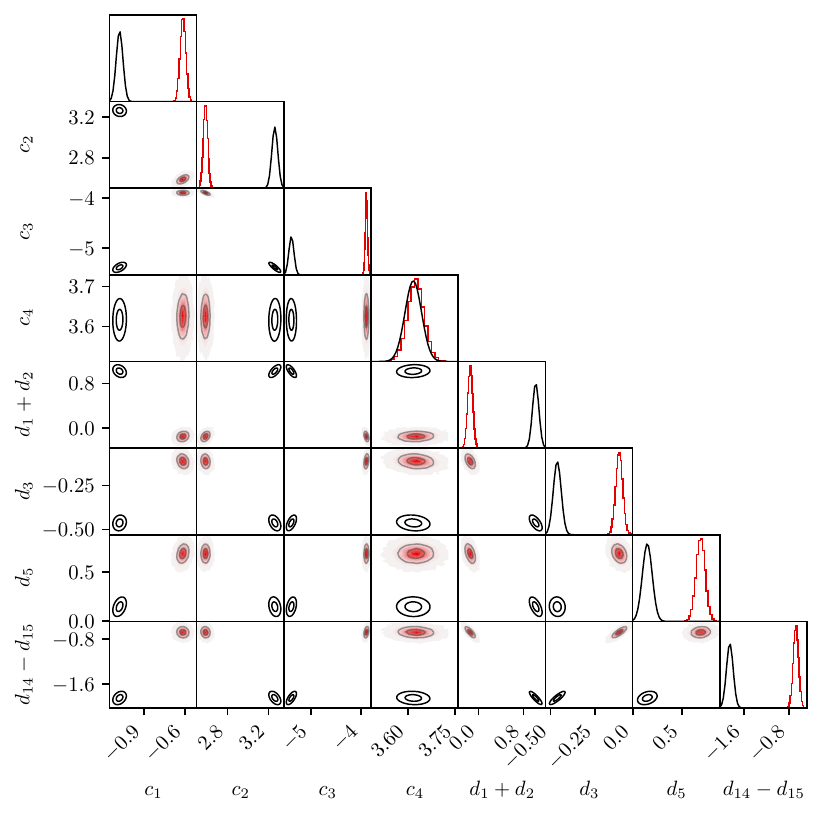}
\caption{Prior and posterior \pdf s for the $c$ and $d$ \piN{} LECs, in units of GeV$^{-1}$ ($c$) and GeV$^{-2}$ ($d$). The posterior (red) was obtained using HMC. The inner (outer) ellipses enclose approximately 39\% (86\%) of the probability mass. Black lines indicate the prior as provided by the Roy-Steiner analysis~\cite{Siemens:2016jwj}.}
\label{fig:piN_contours_n3lo}
\end{figure*}

Figures~\ref{fig:posterior_nlo} and \ref{fig:posterior_nnlo} show corner plots, i.e., marginal uni- and bi-variate \pdf s, for the LEC posterior at \nlo\ and \nnlo, respectively. A subset of the marginal posteriors at \nnnlo\ is shown in Fig.~\ref{fig:posterior_n3lo_extract}. The corner plot of the 31-dimensional \nnnlo\ posterior is too large to print and we thus provide it as Supplemental Material to this paper~\cite{suppl}. In Fig.~\ref{fig:piN_contours_n3lo} we show a comparison of the \piN{} prior and posterior at \nnnlo.

For convenience we list the LECs inferred at each order here, sorted as they appear in the corner plots. The inferred LECs at \nlo\ are
\begin{equation}
\begin{split}
\lecs = (
\Ctnn,
\Ctnp,
\Ctpp,
\Ct_{3S1},
C_{1S0},
C_{3P0},
\\
C_{1P1},
C_{3P1},
C_{3S1},
C_{3S1-3D1},
C_{3P2}).
\end{split}
\end{equation}

The inferred LECs at \nnlo\ are
\begin{equation}
\begin{split}
\lecs = (
c_1,
c_3,
c_4,
\Ctnn,
\Ctnp,
\Ctpp,
\Ct_{3S1},
C_{1S0},
\\
C_{3P0},
C_{1P1},
C_{3P1},
C_{3S1},
C_{3S1-3D1},
C_{3P2}).
\end{split}
\end{equation}

The inferred LECs at \nnnlo\ are
\begin{equation}
\begin{split}
\lecs = (
c_1,
c_2,
c_3,
c_4,
d_1+d_2,
d_3,
d_5,
d_{14} - d_{15},
\\
\Ct_{1S0}^{nn},
\Ct_{1S0}^{np},
\Ct_{1S0}^{pp},
C_{1S0},
D_{1S0},
\Ct_{3S1},
\\
C_{3S1},
C_{3S1-3D1},
D_{3S1},
D_{3S1-3D1},
D_{3D1},
\\
C_{3P0},
D_{3P0},
D_{3P1},
C_{3P1},
C_{3P2},
D_{3P2},
\\
D_{3P2-3F2},
C_{1P1},
D_{1P1},
D_{1D2},
D_{3D2},
D_{3D3}
).
\end{split}
\end{equation}


\section{Multivariate normal approximations of the LEC posteriors}
The LEC posteriors presented in this paper are stored as arrays of samples from which subsequent results are produced. The posteriors are, however, approximately Gaussian, and we therefore provide multivariate normal approximations for convenience. These approximations are available in text format in the supplemental material~\cite{suppl}. The filenames reflect the chiral order, and the contents are structured as follows:
\begin{itemize}
\item The first line shows the ordering of the LECs.
\item Then follows a vector $\vec{\mu}$ of corresponding mean values.
\item Finally, a covariance matrix $\vec{\Sigma}$ encodes the uncertainties.
\end{itemize}
Put together, these quantities define a multivariate normal distribution $\normal(\vec{\mu}, \vec{\Sigma})$ that approximates the LEC posterior.

\section{Posterior predictive distributions: incorporating the $c_4^{pp}$ data}
\label{app:ppd_all}
\begin{figure*}[ht!]
\centering
\includegraphics[width=1.0\linewidth]{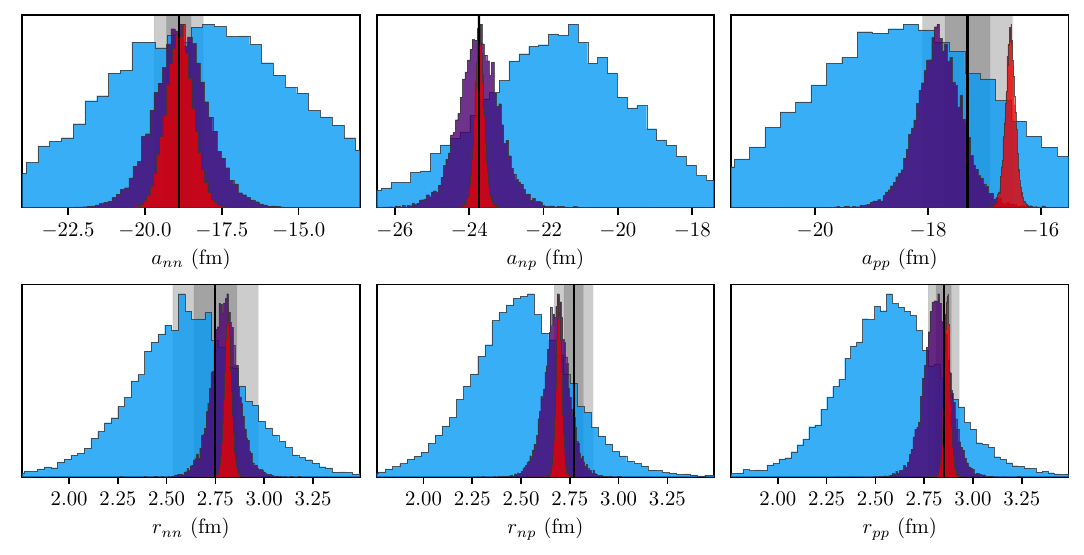}
\caption{\ppd s of scattering lengths and effective ranges at \nlo\ (blue), \nnlo\ (purple), and \nnnlo\ (red) when including all \np\ and \pp\ scattering lengths up to \nnnlo\ to infer $\cbar$. Empirical results are shown as black lines, with corresponding $1\sigma$ $(2\sigma)$ uncertainties as a dark (light) gray area. The empirical results are without electromagnetic effects and gathered from Ref.~\cite{Machleidt:2011zz}, except for $a\subnn$ which we take from Ref.~\cite{Gardestig:2009ya}. Note that the histograms have been scaled such that they have the same peak height.}
\label{fig:ppd_panel_all_eft_data}
\end{figure*}

For completeness, in Fig.~\ref{fig:ppd_panel_all_eft_data} we show the \ppd s for the effective range expansion with an EFT truncation error informed also by the somewhat irregular \nnlo-\nnnlo\ shift in $a\subpp$; see Table~\ref{tab:ere_alpha_star}. This irregularity is however modulated by our prior expectations on $\cbar$ and the convergence of the EFT expansion. In the end, although the value for $\cbar$ increases some, we find that the experimental and theoretical errors are roughly equally sized at \nnlo\, while the theoretical (experimental) error dominates at NLO (N3LO), as before. As a result, mainly the \nlo\ predictions for the scattering length is modified compared to the result in Fig.~\ref{fig:ppd_panel}.

\end{document}